\documentclass[useAMS,usenatbib]{mn2e}
\usepackage{graphicx}
\usepackage[]{natbib}
  \title{The orbital period of Swift~J1816.7-1613 revealed by the Swift-BAT telescope.}

\author[V.\ La Parola et al.]{V.\ La Parola$^{1}$,  A.\ Segreto $^{1}$ , G.\ Cusumano$^{1}$, N.\ Masetti
$^{2}$, A.\ D'A\`i, A.\ Melandri$^{3}$ \\
$^{1}$INAF - Istituto di Astrofisica Spaziale e Fisica Cosmica, Via U.\ La Malfa 153, I-90146 Palermo, Italy\\
$^{2}$INAF - Istituto di Astrofisica Spaziale e Fisica Cosmica di Bologna, via Gobetti 101, 40129, Bologna, Italy\\
$^{3}$  INAF - Brera Astronomical Observatory, via Bianchi 46, 23807, Merate (LC), Italy
\\
}

\begin{document}
\date{}

\pagerange{\pageref{firstpage}--\pageref{lastpage}} \pubyear{2014}

\maketitle
            
\label{firstpage}
\begin{abstract}
We have analyzed the Swift data relevant to the high mass X-ray binary Swift~J1816.7-1613.
The timing analysis of the BAT survey data unveiled a modulation at a period
of P$_0=118.5\pm 0.8$ days that we interpret as the orbital period of the X-ray binary system. 
The modulation is due to a sequence of bright flares, lasting $\sim30$ d, separated by long
quiescence intervals. This behavior is suggestive of a Be binary system, where
periodic or quasi-periodic outbursts are the consequence of 
an enhancement of the accretion flow from the companion star at the periastron passage.
The position of Swift~J1816.7-1613  on the Corbet diagram strengthens this hypothesis.
The broad band 0.2--150 keV spectrum is well modeled with a strongly absorbed
power-law with a flat photon index $\Gamma\sim 0.2$ and 
a cut-off at  $\sim 10$ keV.

\end{abstract}
\begin{keywords}
X-rays: binaries -- X-rays: individual: Swift J1816.7-1613. 

\noindent
Facility: {\it Swift}

\end{keywords}
\section{Introduction}\label{intro}

A significant fraction of cosmic X-ray sources are transient systems, with long
periods of quiescence interrupted by brief intervals of intense emission. 
Large field of view X-ray detectors play a fundamental role to detect and study these
systems. Within this class of telescopes, the Burst  Alert Telescope (BAT,
\citealp{bat}) on board the Swift satellite \citep{swift} is proving to be a valuable
instrument in fulfilling this aim. Since November 2004 BAT has been performing a 
continuous monitoring of the hard X-ray sky and, while it is hunting for gamma-ray bursts, 
it records the flux
variability of known X-ray sources and discovers many new X-ray transients \citep{krimm13}.

In this Letter we analyze the soft and hard X-ray data collected by Swift on 
Swift~J1816.7-1613, a Galactic X-ray transient unveiled by BAT \citep{krimm08}. 
The source was first detected at $\sim 24$ mCrab in the 15-50 keV band
on 2008 March 24 and its intensity rose to $\sim 35$ mCrab on March 29.
The source was also revealed by BAT during two other less-intense outbursts, one
from 2009 July 21 to 2009 August 10 and the other one
from 2011 June 18  to 2011 July 8. During both episodes the
outburst peaked at $\sim30$ mCrab in the 15--50 keV band, \citep{krimm13}.
From the examination of the long term BAT light curve, and following a more 
recent outburst in June 2014, \citet{corbet14} suggest an
orbital periodicity of $\sim151$ d and argue that the sporadic appearance
of the outbursts along the light curve is consistent with the source being a
Be/neutron star binary system. 
The X-ray transient was detected serendipitously in a Chandra ACIS observation 
performed on 2007 February 11 \citep{Halpern08}.
The position of the X-ray source measured by Chandra is  RA=18h 16m 42.66s, 
Dec=-16$^{\circ}$ 13' 23.4'' (J2000). The energy spectrum was modeled by a power law of photon 
index $ \Gamma = 1.2$, a column density N$_{\textrm H}=12 \times 10^{22} \rm cm^{-2}$ and 
a 2--10 keV flux of $\rm 4\times 10^{-12} erg~cm^{-2} s^{-1}$. A timing analysis 
of the Chandra event arrival times has also revealed that the source is 
a high mass X-ray pulsar with a spin period of $\sim 143$ s \citep{Halpern08}. 
The analysis of two follow-up observations performed with the RXTE PCA on 2008 March 29 and 
April 7 confirmed the  pulse period and revealed  weak evidence for a spin-up between
the two RXTE observations with a  $\rm \dot P =-5.93 \times 10^{-7} s~s^{-1}$ \citep{krimm13}. 
Archival data of XMM-Newton  and BeppoSAX-PDS also showed the presence of the source 
at a flux of $\rm 7\times10^{-13} erg~cm^{-2} s^{-1}$  in 2--10 keV
on 2003 March 8 \citep{Halpern08} and  at a flux of $\rm 3.6 \times 10^{-11} 
erg~cm^{-2} s^{-1}$ in 15--30 keV on 1998  September 29 \citep{orlandini08}, respectively.
The companion star has not been identified yet; archival searches in the Chandra region of
the X-ray source were unable to reveal any optical/NIR counterpart \citep{krimm13}. 
 
This Letter is organized as follows. Section 2 describes the Swift data
reduction. Section 3 reports on the timing analysis.
Section 4 describes the broad band spectral analysis. In Section 5 we briefly
discuss our results.

\begin{figure*}
\begin{center}
\centerline{\includegraphics[width=18cm,angle=0]{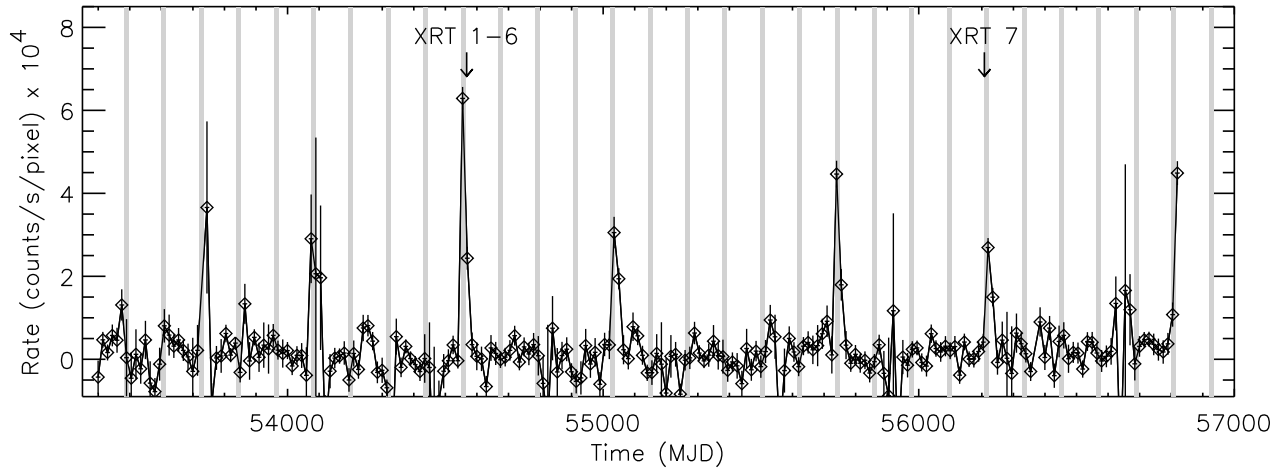}}
\caption[]
{BAT light curve in the 15-45 keV band. The bin length corresponds to a time interval of  
15\,days.  The vertical shaded areas 
are separated by 118.5 days and are in phase with the peak position of the brightest outburst at
MJD 54556.1. 
               }               
                \label{lc} 
        \end{center}
        \end{figure*}

\section{Observations and data reduction}\label{data}

The Swift-BAT survey raw data collected between December 2004 and June 2014
were processed with dedicated software \citep{segreto10} 
that computes all-sky maps in several energy bands between 15 and 150 keV,
performs source detection on these maps and for each detected source
produces standard products such as light curves and spectra.
Swift~J1816.7-1613 was inside the BAT field of view 
for a total of 38.1 Ms. The source was detected in the 15--150 keV all-sky map 
with a signal to noise ratio of 10.1 standard 
deviations and of 12.8  standard 
deviations in the 15--45 keV all-sky map, where its signal-to-noise
is maximized. The latter energy band was used to extract
the light curve with the maximum time resolution ($\sim300$ s) allowed by the
Swift-BAT survey data.  
The spectrum is extracted in 8 energy channels and analyzed 
using the official BAT spectral redistribution matrix
\footnote{http://heasarc.gsfc.nasa.gov/docs/heasarc/caldb/data/swift/bat/index.html}.

\begin{table*}
\caption{XRT observations log. The quoted orbital phase refers to the 
profile reported in the lower panel of Figure~\ref{period-best}b. \label{log}}
\scriptsize
\begin{center}
\begin{tabular}{r l l l l l l l} \hline
Obs \# & Obs ID     & $T_{start}$ & $T_{elapsed}$ & Exposure & Rate            & Orb.  Phase & Spin Period  \\
       &            &  MJD        &  (s)          & (s)      &  (c/s)              &         & (s)          \\ \hline \hline
1      &00031187001 & 54557.783   &  1981.6       &  1945.4  &$0.32\pm0.01$        &0.44     & $143.3 \pm 1.3$  \\
2      &00031188001 & 54560.210   & 12490.3       &  1927.9  &$0.42\pm0.02$        &0.46     & $142.9 \pm 2.5$   \\
3      &00031188002 & 54562.547   & 18571.9       & 1932.9   &$0.28\pm0.01$        &0.48     & $143.4 \pm 5.6$   \\
4      &00031188003 & 54574.800   & 12270.3       & 1212.9   &$0.064\pm0.08$       &0.58     &  ---            \\
5      &00031188004 & 54576.413   & 12170.6       & 1251.1   &$0.068\pm0.008$      &0.60     &  ---             \\ 
6      &00031188005 & 54578.218   & 12115.8       & 1935.4   &$0.034\pm0.005$      &0.61     &  ---             \\ 
7      &00044103001 & 56207.983   & 489.7         &  486.8   &$0.07\pm0.01$        &0.37     &  ---            \\ \hline
\end{tabular}
\end{center}
\end{table*}

Swift-XRT \citep{xrt} observed Swift~J1816.7-1613  seven times. The source was observed 
in Photon Counting (PC) mode  \citep{hill04}
in all the  observations, for a total exposure time of $\sim10.7$ ks. 
The details on the Swift-XRT observations are reported in Table~\ref{log}.
The XRT data were processed with standard procedures ({\sc xrtpipeline} 
v.0.12.8) using the {\sc ftools} in the {\sc heasoft} package (v 6.15) and the 
products were extracted adopting a grade filtering of 0--12. The source events 
were extracted from a circular region of 20 pixel radius (1 pixel = 2.36'') 
centered on the Chandra source position, while 
the background for spectral analysis was extracted from an annular region 
centered on the source, with an inner radius of 70 pixels and an outer radius of
130 pixels, to avoid  the contamination due to the PSF tail
of Swift~J1816.7-1613. All source event arrival times were converted to 
the Solar System barycentre with the task 
{\sc barycorr}\footnote{http://http://heasarc.gsfc.nasa.gov/ftools/caldb/help/barycorr.html}.
XRT ancillary response files for each observation were generated 
with {\sc xrtmkarf}\footnote{http://heasarc.gsfc.nasa.gov/ftools/caldb/help/xrtmkarf.html}. 
Besides spectra for each single XRT observation, we also extracted a cumulative source and
background spectrum from observations 1 to 3, excluding observations 4 to 7 because of
their low statistical content.
The ancillary files relevant to each XRT observation were 
combined into a single one using {\sc addarf} that weights them by the exposure times of the 
relevant source spectra. 
Finally, all the spectra were re-binned with a 
minimum of 20 counts per energy channel, in order to allow the use of the 
$\chi^2$ statistics. We used the spectral redistribution matrix v014 and the 
spectral analysis was performed using {\sc xspec} v.12.5.  
Errors are at 90\,\% confidence level for a single parameter, if not stated otherwise.

\section{Timing analysis and results}\label{timing} 
Figure~\ref{lc} shows the light curve of Swift~J1816.7-1613 in the 15-45 keV band, between
November 2004 and June 2014 (with a bin of length 15 days), where
several outburst episodes of similar duration ($\sim 30$ days) and intensity are visible. 
This behaviour is
suggestive of a Be/X-ray binary system (as already suggested by \citealp{corbet14}) where the
appearance of type I outbursts is related to the periastron passage of the compact object.  
The similarity of all the observed flares both in duration and intensity suggests that none of
them is a Type II outburst uncorrelated with the orbital modulation.
The outbursts peak times are MJD $53745\pm30$, $54088\pm9$, $54556.1\pm1.0$, $55041.1\pm1.0$, 
$55742.6\pm1.5$, $56224.5\pm1.0$, $56816.4\pm1.0$ , where the peak time and its error are 
evaluated fitting a Gaussian to the relevant portion of the light curve (produced with a bin 
time of 5 days).  The larger uncertainty in the first and second outbursts are due to the 
fact that they are detected with lower significance, because they are
covered only partially by the survey data.
According to the sequence of outbursts observed during our monitoring, the periodicity, if any, 
is constrained to be a submultiple of the distance between any pair of consecutive outbursts. This
is evident when we observe that the time interval between the second pair of outbursts ($\sim 470$
d) is not a multiple of the time interval between the first two outburst ($\sim 340$ d, the
shortest interval in the sequence). 

\begin{figure}
\begin{center}
\centerline{\includegraphics[width=8cm]{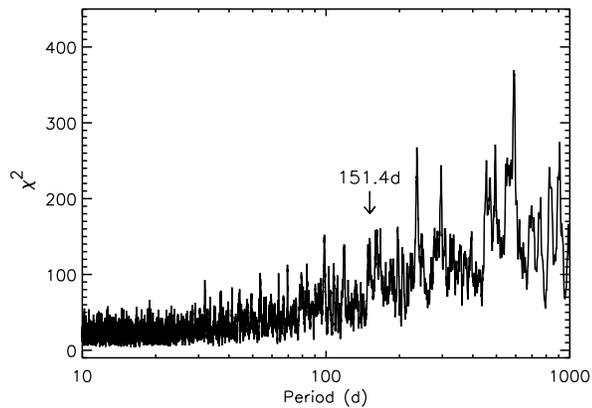}}
\caption[]{Periodogram of {\it Swift}-BAT (15--45\,keV) data for
Swift~J1816.7-1613. The arrow marks the period suggested by \citet{corbet14}.
                \label{totalp}
}
\end{center}
\end{figure}

In order to verify if the appearance of these outbursts has
a periodic behaviour we have performed a timing analysis on this light curve using the
folding technique \citep{leahy83}. 
This technique consists in the production of a count rate profile for a set of trial periods  
by folding the count rates in N phase bins and in the evaluation of 
the $\chi^2$ value with respect to the average count rate for the profile corresponding  
to each trial period. A large value of  
$\chi^2$ represents a clue of a periodic modulation.
The periodogram has been produced in the 1--1000 d period range 
with a step of P$^{2}/(N  \,\Delta$T$_{\rm BAT})$ (P is 
the trial period, $N=16$ is the number of phase bins used to build the profile 
and $\Delta$T$_{\rm BAT}\sim 298.2$ Ms is the  data time span). This periodogram (Figure~\ref{totalp}) 
is quite noisy for test periods higher than $\sim 100$ d. A few peaks emerge over the noise, however none
of them has enough power to be claimed as a strong indication of periodicity, nor is consistent either with
151.4 d \citep{corbet14} or with being a submultiple of both 340 d and 470 d. 
We have then repeated the folding analysis after selecting from
the light curve  only 200-day data intervals centered on the observed outbursts: this selection
allows us to exclude light curve intervals that would introduce only noise in the following timing
analysis, and still explore the possible range of outburst periodicities.  

\begin{figure}
\begin{center}
\centerline{\includegraphics[width=8cm]{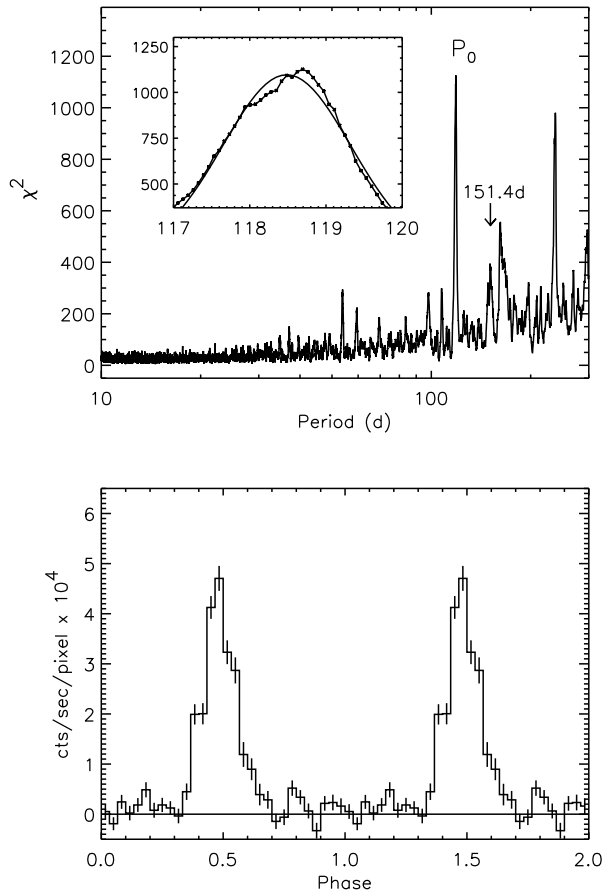}}
\caption[]{{\bf Top panel}: Periodogram obtained selecting only 200-day time 
intervals centered on the observed outbursts; the period suggested by \citet{corbet14} is marked with an
arrow. The inset shows a close-up view of the periodogram
around $\rm P_0=118.5$ d and the best-fitting Gaussian function to the peak.
{\bf Bottom panel}:  Swift BAT light curve folded with a period of $\rm P_0$  in 30 phase 
bin selecting only 200-day time intervals centered on the observed flares.

                \label{period-best}
}
\end{center}
\end{figure}

Figure~\ref{period-best} (top panel) shows the
periodogram obtained from the selected data, where a prominent feature emerges at $\sim118$ d. 
Its peak can be fitted with a Gaussian function (Fig~\ref{period-best} - inset in top panel), 
obtaining a peak centroid at $\rm P_0=118.5\pm0.8$ d, where the error is the standard deviation of 
the best-fitting Gaussian.  This value is consistent with being a submultiple of both the two shortest
time intervals between the outbursts ($\sim 340$ d and $\sim 470$ d). 

The bottom panel of Figure~\ref{period-best} shows the pulse profile obtained by folding the data 
in 30 phase bins with a periodicity of $\rm P_0$ and 
T$_{\rm epoch}= \rm MJD~55216.6$. The profile shows a narrow peak emerging by at
least two orders of
magnitude over a flat plateau whose intensity is marginally above zero. The peak 
is at phase $0.48\pm 0.5$ and 
corresponds to  MJD $(55273.7\pm 2)\pm  n \times $P$_0$, where
the  uncertainty is half phase bin.

Table~\ref{log} reports the average count rate recorded during the XRT observations and the 
relevant orbital phase evaluated with respect to P$_0$ and T$_{\rm epoch}$. 
Observations 1-3  performed by XRT on Swift~J1816.7-1613 fall in correspondence
of the flare peaking at MJD 54556 in Fig.~\ref{lc}, Obs. 4-6 corresponds to the end tail 
of the same flare, while Obs. 7 is just before the flare peaking at MJD 56224.

We performed a timing analysis on the XRT data searching for the presence of
the periodic modulation detected by Chandra \citep{Halpern08} and confirmed by RXTE
\citep{krimm13}.
XRT observations are fragmented into snapshots  of different duration and
time separation. 
This causes spurious features to emerge when the timing analysis is
performed on data belonging to more than one snapshot, and affects 
the detection of a source periodic signal.
To avoid these systematics we have performed a folding analysis separately on each snapshot 
searching in a period range 100--200 s. The periodic modulation has been revealed
in the periodograms derived from the snapshots of Obs 1, 2 and 3 (Table 1); the statistics 
content in the other XRT observations is
too poor to allow for a periodic modulation to emerge over the noise.
In each snapshot periodogram a period was 
associated to the centroid of the Gaussian that best fit the revealed feature 
while the error is evaluated applying equation 6.a in \citet{leahy87}.
The period relevant to each observation was then  evaluated by weighting the periods
relevant to each snapshot with  the  inverse  square  of  their errors.
Table ~\ref{log} lists the resulting periods.
The period averaged among  Obs 1, 2 and
3 is $143.2 \pm  1.1$ s.

\section{Spectral Analysis} \label{spectral}
As shown in Fig.~\ref{lc}, BAT observed several outburst episodes along its monitoring 
of Swift~J1816.7-1613.
Selecting only the data in these intervals (30-day time intervals centered on the outbursts peak
times), the source is detected at a significance
level of about 30 standard deviations while outside these outbursts  Swift~J1816.7-1613 
shows a weak emission below the detection threshold (S/N $\sim 4$). 

We have performed a broad band spectral analysis relevant to the periods of outburst 
emission.
First, we verified that no significant spectral variability is present
among XRT obs. 1-3 and among  the outbursts observed by BAT;
fitting each XRT spectrum  with an absorbed power-law we found that both the best fit 
absorption columns and photon indeces were consistent within the statistical errors.  
The BAT spectra relevant to each flare episode, fitted with a simple power-law,
also showed no significant spectral variation. 
We therefore summed the XRT data of observation 1-3 and the BAT data collected during 
all the outburst episodes and combined  the resulting XRT and  BAT spectra to 
perform a broad band spectral analysis.
Trial models were multiplied by a constant to account for any 
inter-calibration uncertainty between the two telescopes and for different flux 
levels among the two spectra. This constant was frozen to unity for the XRT spectrum
 and was left free to vary for the BAT spectrum.  

An absorbed power-law model ({\tt cons*phabs*powerlaw}) was rejected because of its 
unacceptable $\chi^2$ of 221 over 83 degrees of freedom (dof).  
The fit is significantly  improved by including a cutoff in the above model 
({\tt cons*phabs*cutoffpl}), with a resulting $\chi^2$ of 92 over 82 dof, and an F-test
probability of $\sim3\times10^{-4}$ of obtaining this improvement by chance. 

\begin{figure}
\begin{center}
\centerline{\includegraphics[width=5.5cm,angle=-90]{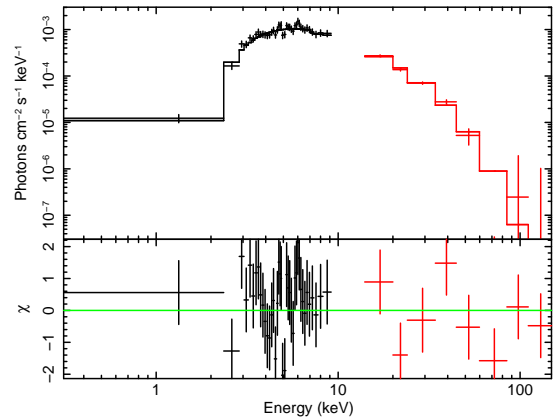}}
\caption[]
{Swift~J1816.7-1613 broad band spectrum collected during the outbursts. {\bf Top panel}:
XRT and BAT unfolded spectrum and best fit {\tt phabs*(cutoffpl)} model. 
{\bf Bottom panel}: Residuals in unit of standard deviations for the 
{\tt phabs*(cutoffpl)} model.
               }
                \label{spec}
        \end{center}
        \end{figure}

Figure~\ref{spec} shows the data and best fit model and the residuals, while  
table \ref{fit} reports the best-fitting parameters.

\begin{table}
\begin{tabular}{ r l l}
\hline
Parameter & Best fit value & Units    \\ \hline \hline
N$_{\textrm H}$   & $10.2^{+1.8}_{-1.6} \times 10^{22}$ & $\rm cm^{-2}$\\
$\Gamma$  &$0.1^{+0.4}_{-0.4}$&          \\
$E_{cut}$ &$9.2^{+1.7}_{-1.3}$ & keV\\
$N$       &$ 2.9^{+2.4}_{-1.3}\times 10^{-3}$ &ph $\rm keV^{-1} cm^{-2} s^{-1}$ at 1 keV \\
$\rm C_{BAT}$&$0.9^{+0.2}_{-0.2}$&\\
$\rm F$ (0.2--10 keV)&$1.0\times 10^{-10}$& erg s$^{-1}$ cm$^{-2}$\\
$\rm F$ (15--150 keV)&$1.3\times 10^{-10}$& erg s$^{-1}$ cm$^{-2}$\\
$\chi^2$   &92 (82 d.o.f.) & \\ \hline
\end{tabular}
\caption{Best fit spectral parameters.  $\rm C_{BAT}$ is the
constant factor to be multiplied to the model in order to match the XRT and BAT data.
We report unabsorbed fluxes for the standard XRT (0.2--10 keV) and BAT (15--150 keV)
energy bands. \label{fit}}
\end{table}

\section{Discussion and conclusions}\label{conclusion}

\begin{figure}
\begin{center}
\centerline{\includegraphics[width=8cm]{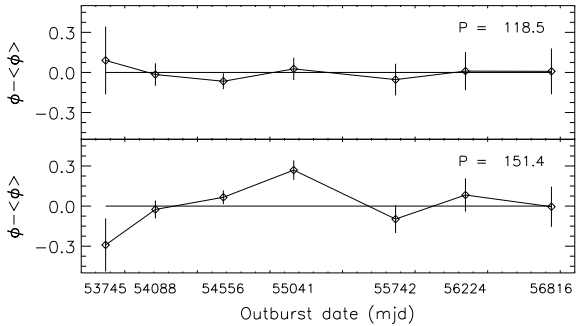}}
\caption[]{Residual phase of the source outbursts with respect to the  periodicity of 118.5 d 
(upper panel), and to the periodicity suggested in \citet{corbet14} (lower panel).
                \label{chip}
}
\end{center}
\end{figure}

The light curve of Swift~J1816.7-1613 is characterized by a sequence of very bright outbursts
lasting $\sim 30$ d, and long quiescent time intervals. During the 113 months of BAT survey 
monitoring we observe seven outbursts, that we find to have a periodicity  P$_0= 118.5\pm0.8$ d,
that we interpret as the orbital periodicity of the source. 
Our result is at odds with the periodicity of $\sim 151$ d obtained by \citet{corbet14}. To
compare the two results we have analysed the phase $\phi$ of each of the observed outbursts with 
respect to 
P$_0= 118.5\pm0.8$ d and to $\rm P=151.4\pm 1$ d, and plotted the residual phase 
$\phi$-$<\phi>$, where $<\phi>$ is the average phase value for the set of observed outburst with
respect to each of the two periods (Figure~\ref{chip}). The latter values are 
expected to be all consistent with 0 if the test periodicity describes well the outburst 
sequence.  If we apply a $\chi^2$ test to the residual phases,
assuming an expected value of 0, we find a reduced  $\chi^2$ of 0.5 for P=118.5 days and 3.1 for
P=151.4 d, respectively. The higher $\chi^2$ obtained for P=151.4 d is due mainly to  the
outburst at MJD 55041  that is not consistent with this periodicity.

\begin{figure}
\begin{center}
\centerline{\includegraphics[width=8cm,angle=0]{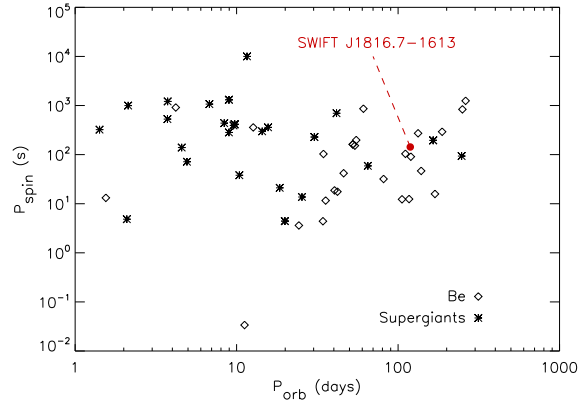}}
\caption[Corbet diagram]{
Corbet diagram for HMXBs with known spin and orbital period. Diamond and star points
represent the Be and supergiant systems, respectively. The red filled circle
marks the position of Swift~J1816.7-1613.
                }
                \label{corbet}
        \end{center}
        \end{figure}

The BAT light curve folded at P$_0$ is characterized by a narrow peak emerging 
over a flat plateau, where the off-peak emission is at least two order of magnitude lower than the
peak intensity. The peak  has a duration of 
$\sim 30\%$ of P$_0$, corrsponding to $\sim 30$ days. 
Thus the emission of this source is concentrated in a short fraction of its
orbit. Moreover, the outbursts do not occur at every cycle: in the sequence 
observed during our monitoring they are separated by 3 to 6 cycles.
This behaviour is commonly observed in X-ray binaries with a Be star as a companion: 
these stars are likely characterized by an equatorial extended disc, fed from
material ejected from the star itself because of its rapid spin rotation (see e.g. 
\citealp{reig11} for a recent review). In such binary 
systems, the stellar disk is often 
oriented on a plane different from the orbital plane of the neutron star. 
Periodic or quasi-periodic outbursts, lasting a fraction of 10-30\% 
of the orbital period, 
peaking close to the periastron passage of the neutron star are the consequence of 
an enhancement of the accretion flow from the companion star as it passes  
close to the circumstellar disk.
The position of Swift~J1816.7-1613  on the Corbet diagram \citep{corbet86} in Figure~\ref{corbet} 
shows that the source lies in the Be transients region, 
adding a further strong hint on its nature as a Be/X-ray binary system.

A timing analysis on the XRT data confirms the presence of
the periodic modulation first revealed by Chandra \citep{Halpern08} and confirmed by RXTE
follow-up observations \citep{krimm13}.  The spin periodicity, detected in the XRT  Obs 1, 2 and
3 is $143.2 \pm  1.1$ s. 


Finally, the broad band 0.2--150 keV spectrum collected during the
outbursts is modeled with an absorbed
power-law with a flat photon index $\Gamma\sim 0.1$ and a steepening in the BAT energy range
modeled with a cut-off at an energy of $\sim 10$ keV, and an average flux of 
$\rm (1.30\pm 0.08)\times 10^{-10}~erg~cm^{-2}s^{-1}$. The spectrum is strongly absorbed, with
column density one order of magnitude higher than the average value along the line of sight.
Assuming this same spectral shape, the off-outburst average flux is 
$\rm (2.9\pm1.1)\times 10^{-12}~erg~cm^{-2}s^{-1}$.

\section*{Acknowledgments}
This work has been supported
by ASI grant I/011/07/0.

\bibliographystyle{aa}

\begin{thebibliography}{}


\bibitem[Barthelmy et al.(2005)]{bat} Barthelmy, S.~D., et 
al.\ 2005, Space Science Reviews, 120, 143 



\bibitem[Burrows et al.(2004)]{xrt} Burrows, D.~N., Hill, 
J.~E., Nousek, J.~A., et al.\ 2004, SPIE, 5165, 201 


\bibitem[Corbet(1986)]{corbet86} Corbet, R.~H.~D., 1986, MNRAS, 220, 1047

\bibitem[Corbet \& Krimm(2014)]{corbet14} Corbet, R.~H.~D., \& Krimm, H.~A., 2014, ATEL 6253

\bibitem[Dickey \& Lockman(1990)]{dickey90} Dickey, J.~M., \& Lockman, F.~J.\ 
1990, ARAA, 28, 215 

\bibitem[Gehrels et al.(2004)]{swift} Gehrels, N., et al.\ 
2004, ApJ, 611, 1005 


\bibitem[Halpern \& Gotthelf(2008)]{Halpern08} Halpern, J.~P., \& Gotthelf, E.~V.\ 2008, The
Astronomer's Telegram, 1457, 1 

\bibitem[Hill et al.(2004)]{hill04} Hill, J.~E., Burrows, 
D.~N., Nousek, J.~A., et al.\ 2004, SPIE, 5165, 217 

\bibitem[Kennea et al.(2005)]{atel459} Kennea J. A., Burrows D. N., Nousek
J. A., Chester M., Roming P., Barthelmy S., Gehrels N., Beckmann V., Soldi S.,
2005, ATEL 459

\bibitem[Krimm et al.(2008)]{krimm08} Krimm, H.~A., Barthelmy, 
S.~D., Baumgartner, W., et al.\ 2008, The Astronomer's Telegram, 1456, 1 


\bibitem[Krimm et al.(2013)]{krimm13} Krimm, H.~A., Holland, 
S.~T., Corbet, R.~H.~D., et al.\ 2013, ApJS, 209, 14 

\bibitem[Leahy(1987)]{leahy87} Leahy, D.~A., \ 1987, A\&A, 180, 275 

\bibitem[Leahy et al.(1983)]{leahy83} Leahy, D.~A., Elsner, 
R.~F., \& Weisskopf, M.~C.\ 1983, ApJ, 272, 256 

\bibitem[Orlandini \& Frontera(2008)]{orlandini08} Orlandini, M., \& Frontera, F.\ 2008, The
Astronomer's Telegram, 1462, 1 

\bibitem[Reig(2011)]{reig11} Reig, P.\ 2011, Ap\&SS, 332, 1 

\bibitem[Segreto et al.(2010)]{segreto10} Segreto, A., Cusumano, G., Ferrigno, C., La Parola, 
V., Mangano, V., Mineo, T., \& Romano, P.\ 2010, A\&A, 510, A47 


\end{thebibliography}

{}
\label{lastpage}
\end{document}